%% file: paper-arxiv-ml-artificial-historic.tex
\documentclass{article}

\input{./shared/arxivSettings}

\title{Prediction of `artificial' urban archetypes at the pedestrian-scale through a synthesis of domain expertise with machine learning methods}

\date{} 

\input{./shared/author.tex}

\renewcommand{\shorttitle}{Prediction of artificial urban archetypes}

\hypersetup{
	pdftitle={
		Prediction of `artificial' urban archetypes at the pedestrian-scale through a synthesis of domain expertise with machine learning methods
	},
	pdfsubject={
		physics.soc-ph
	},
	pdfauthor={
		Gareth~D.~Simons,
	},
	pdfkeywords={
		computation,
		data science,
		land-use analysis,
		machine learning,
		morphometrics,
		network analysis,
		spatial analysis,
		unsupervised ML,
		urban analytics,
		urban planning,
		urban morphology,
		urbanism
	},
}

\begin{document}
\maketitle
\begin{abstract}
	\input{./content/0_abstract.tex}
\end{abstract}
\input{./content/1_ml_predictions.tex}
\input{./content/2_new_towns_info.tex}
\input{./content/3_towns_dataset.tex}
\input{./content/4_new_towns_classifiers.tex}

\input{./content/5_summary.tex}

\setcounter{figure}{0}
\makeatletter 
\renewcommand{\thefigure}{S\@arabic\c@figure}
\makeatother

\setcounter{table}{0}
\makeatletter 
\renewcommand{\thetable}{S\@arabic\c@table}
\makeatother

\section{Supplementary Material}
\input{./content/S1.tex}

\clearpage
\section{Citations}
\printbibliography[heading=none]{}

\section{Acknowledgements}
\input{./shared/acknowledge_phd.tex}
\input{./shared/acknowledge_data.tex}

\end{document}

%% file: shared/arxivSettings.tex
\usepackage{./shared/arxiv}
\usepackage[natbib=true,style=authoryear,doi=true,isbn=true,url=false,eprint=true]{biblatex}

\usepackage[utf8]{inputenc} 
\usepackage[T1]{fontenc}    
\usepackage{hyperref}       
\usepackage{url}            
\usepackage{booktabs}       
\usepackage{amsfonts}       
\usepackage{nicefrac}       
\usepackage{microtype}      
\usepackage{graphicx}

\usepackage{doi}

\usepackage{gensymb} 
\usepackage{rotating} 
\usepackage{caption}
\usepackage{subcaption}
\usepackage[nobottomtitles*]{titlesec}
\usepackage{amsmath}
\usepackage{setspace}

\newcommand{\code}[1]{\lstinline[language=bash, basicstyle=\ttfamily\small]|#1|}
\usepackage{color}
\definecolor{lightgrey}{rgb}{0.975, 0.975, 0.975}
\definecolor{midgrey}{rgb}{0.6, 0.6, 0.6}
\definecolor{deepgrey}{rgb}{0.2, 0.2, 0.2}
\definecolor{codegreen}{rgb}{0, 0.7, 0}
\definecolor{codepink}{rgb}{0.8196, 0.10, 0.654}
\usepackage{listings}

%% file: shared/author.tex
\author{
	\href{https://orcid.org/0000-0003-3790-0638}{
		\includegraphics[width=0.25cm, height=0.25cm, keepaspectratio]{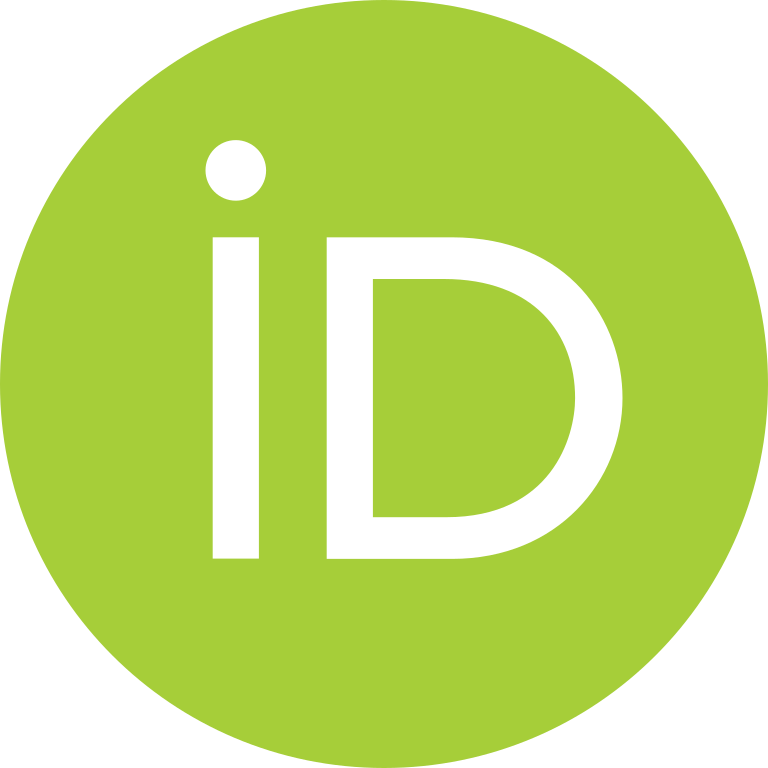}
		\hspace{1mm}Gareth D. Simons
	}
	\thanks{Benchmark Urbanism \texttt{gareth@benchmarkurbanism.com}}
}

%% file: content/0_abstract.tex
The vitality of urban spaces has been steadily undermined by the pervasive adoption of car-centric forms of urban development as characterised by lower densities, street networks offering poor connectivity for pedestrians, and a lack of accessible land-uses; yet, even if these issues have been clearly framed for some time, the problem persists in new forms of planning. It is here posited that a synthesis of domain knowledge and machine learning methods allows for the creation of robust toolsets against which newly proposed developments can be benchmarked in a more rigorous manner in the interest of greater accountability and better-evidenced decision-making.

A worked example develops a sequence of machine learning models that distinguishing `artificial' towns from their more walkable and mixed-use `historical' equivalents. The dataset is developed from network centrality, mixed-use, land-use accessibility, and population density measures as proxies for spatial complexity, which are computed at the pedestrian-scale for 931 towns and cities in Great Britain. Using officially designated `New Towns' as a departure point, a series of clues is then developed. First, using an iterative human-in-the-loop procedure, a supervised classifier (Extra-Trees) is cultivated from which 185 `artificial' locations are identified based on data aggregated to respective town or city boundaries. This information is then used to train supervised and semi-supervised (M2) deep neural network classifiers against the higher resolution dataset.

The models broadly align with intuitions expressed by urbanists and show potential for continued development to broach ensuing challenges pertaining to: selection of curated training exemplars; further development of techniques to accentuate localised scales of analysis; and methods for the calibration of model probabilities to align with the intuitions of domain experts.

%% file: content/1_ml_predictions.tex
\section{Prediction of urban archetypes with deep neural networks}\label{prediction-of-archetypes}

In 2012, \citet*{Krizhevsky2012} introduced a revolutionary machine learning model: it combined massive datasets, convolutional layers, and deep-learning to attain best-in-class classification accuracies on the ImageNet database, consisting of more than 15 million images in over 22,000 categories. The ideas and techniques were not necessarily new, but the authors noted that the depth of their neural network, in this case, five convolutional layers and three fully connected layers, was pivotal to the model's performance. The subsequent upsurge in the use of ever-larger datasets combined with ever-deeper neural networks has led to machine learning models with human --- or better than human --- levels of performance. Deep learning has attained a near-mythical status and is a prevalent feature in the rapid development of AI, with \citet{Silver2017} going on to claim that AlphaGo, an AI underpinned by deep learning, had learned `superhuman proficiency' in the game of Go from scratch without the aid of human knowledge.

On closer scrutiny, claims that such systems are truly capable of developing human-like intelligence, especially from `tabula rasa', tend to be overstated, and further breakthroughs will be required before truly generalisable artificial intelligence can become a reality \citep{Mitchell2021}. \citet{Marcus2018a} argues that AlphaGo's intelligence is not truly `innate': human knowledge has entered the system in the form of a Monte Carlo tree search algorithm, thus empowering the system with the techniques necessary to learn solutions specific to the challenge at hand. Further, the model's intelligence is not generalisable: learning other games implies retraining, and the model cannot solve broader classes of problems that young children may trivially solve. The tremendous volumes of data and the great difficulty in generalising deep neural nets to other problems draws sharp contrasts to the human mind \citep{Sinz2019} whose innate structures appear to facilitate an ability to form rapid and powerful abstractions that generalise well to varied forms of problem-solving. These challenges underscore an important and oft understated reality: deep learning is a tremendously powerful, but also fickle, tool \citep{Marcus2018}. It is brittle by nature: data-hungry, narrowly focused, and easily fooled. Neural networks may learn patterns but cannot `see the forest for the trees'. If representative patterns are not present in the data or go undetected by a model's structure or \emph{loss} function, the model `does not know what it does not know'. Such models may consequently behave contrary to best intentions by being needlessly complex \citep{Rudin2019}, biased or ignorant of unrepresented or unfairly represented classes within the data \citep{Rudin2019, Corbett-Davies2018}, and are generally difficult to develop or reproduce \citep{Henderson2017}.

The proverbial notion that machine learning is an autonomous technology that can magically conjure meaning out of meaningless jumbles of data and that deep-learning infused AI and robotics technologies will soon usher in a utopian future must therefore invoke cynicism. However, it is also important to note that nascent machine learning methods remain amongst the most powerful and valuable tools currently at the disposal of the scientific community and that many of the perceived shortcomings are attributable to a disconnect between the hype associated with the models and an otherwise more realistic understanding of their nature and limitations. The contributions of humans to model development tends to be understated \citep{Marcus2018a}: for these models to be meaningful and trustworthy, they require large amounts of domain-specific information imparted at various stages of the model's development process. In this sense, ML is a powerful sidekick, but one that is potentially prone to naive assumptions or misbehaviour if left to its own devices. These models require interaction and oversight in a process akin to a `dance with data'. Datasets have to be selected and prepared in a manner that accurately represents the nature of the data that we want the algorithms to learn, targets and loss functions coerce models in the right direction, and regularisation methods and testing procedures are necessary to ensure that models are capable of generalisation to unseen samples in a manner that is realistic and fair for the task at hand.

Urban scientists consequently need to be aware of how datasets, data science methods, and machine learning models may ultimately affect day-to-day decisions and policies \citep{Duarte2020}, and how that misinformed models may end up being used to justify courses of action affecting city-citizens and the urban environment for the worse. There is a danger in chasing misguided accuracy metrics or `buzz-friendly' marketing pitches: models can be accurate, but meaningless. An illustrative example is the application of simple error or accuracy rates to unbalanced datasets. Class imbalances are regularly faced by real-world data analysis situations when labels for one class substantially overpower another's presence, as may be the case with credit card fraud data; the minority class (fraudulent transactions) may be infinitesimally more diminutive than the majority class. When training a classifier against an unbalanced dataset using simple accuracy rates, the algorithm may opt to completely ignore the minority class (e.g. inferring that all credit card transactions are not fraudulent) while claiming an accuracy approaching 100\%. Various strategies exist for the temperance of class imbalance problems\footnote{Examples include: undersampling the majority class, oversampling the minority class, adjusting the costs associated with losses from respective classes \citep{Chawla2004}; use of more nuanced accuracy metrics such as Receiver Operating Characteristic curves (the true positive rate plotted against the false positive rate) or F1 scores (weighted average of precision and recall) \citep{Garcia2009}; and calibration techniques for correcting the distributions of probabilistic classifications \citep{Pozzolo2015}.}. Nevertheless, the application of such techniques requires intervention through the role of an informed data scientist who, in turn, needs to be aware of the potential presence of such imbalances and how overlooking these may have far-reaching ramifications. This example reflects the broader issue: the development of predictive machine learning models may require a substantial degree of nurturing, testing, and oversight to understand how the model `thinks' and `reacts' to the data and to guard against unintended forms of behaviour. In this regard, visualisation methods can be of particular importance because they can helpt to convey how the models work while allowing domain experts, who may not have direct low-level knowledge of how these models function, an opportunity to provide feedback on suspicious forms of behaviour.

Whereas the misuse of data science methods for any variety of problematic workflows or end-purposes exists, these methods also hold the potential for scalable and rigorous forms of sensible analysis if used with sufficient safeguards and rigorous oversight from those with detailed knowledge of the domain of interest. Contrarily, it bears emphasis that throngs of architects, urban designers, planners, engineers, civic officials, and NIMBYs have, in turn, been directly responsible for a trail of ill-conceived urban interventions, and this cannot be blamed on statistics or models so much as a human proclivity towards reductionism and self-interest. Although humans are better than machines at generalising problems, they can also be susceptible to wistful narratives or easily waylaid by idealistic pursuits or profit-driven motives. Further, even where skilled and perceptive urban designers and planners are well-aware of implicit biases underpinning problematic planning proposals, they may be at a loss to bolster better-informed decision-making against hearsay or political pressures. Against this backdrop, an interesting question can be posed: can we connect the strong suits of domain experts, who may intuitively understand the issues at hand, to the strong suits of algorithms capable of exhaustively exploring and laying bare the solution space in a robust and scalable manner? How might tools that synthesise qualitative knowledge with quantitative approaches build an accountable evidence base within the context of politically wrangled decision-making processes?

%% file: content/2_new_towns_info.tex
\section{Historical context of New Towns}\label{new-towns}

Ever since the backlash commenced against (to use Christopher Alexander's term) `artificial' towns and cities \citep{Jacobs1961, Alexander1967}, modernity's failed planning experiments have been easy to reject out-of-hand as unsuccessful and misguided. An endless stream of discourse continues to reinforce this narrative while lamenting the state of artificially planned communities, perhaps best epitomised by discussions framing the broken planning paradigms enshrined by suburban sprawl \citep{Katz1994, Langdon1994, Duany2000, Ellin1999}. Nevertheless, the undermining of the pedestrian and public realm continues unabated \citep{TransportforNewHomes2018} and, as has been argued for Smart Cities, idealised conceptions of urban life re-imagined in the name of engineered efficiencies echo a hauntingly familiar refrain \citep{Greenfield2013, Hill2013, Townsend2013, Sterling2014}.

Whereas these forms of development are doubtlessly problematic, it is also true that they are often guided by good, if waylaid, intentions. From this perspective, historic discussions surrounding the formulation of the New Towns Act of 1946 may appear more nuanced than might be expected. Second World War bombing raids had left untold destruction on London's urban fabric. The magnitude of this damage is hard to comprehend or convey and is perhaps best epitomised by a historic photograph: `St. Paul's Survives', wherein St. Paul's Cathedral --- which by sheer luck had been spared substantial damage --- stands surrounded by destruction and smoke emanating from nearby fires \citep{StPaulsSurvivesWikipedia}. By War's end, bombing damage from air-raids had greatly exacerbated any existent need for housing; meanwhile, Letchworth and Welwyn Garden Cities, for which development had commenced in the decades before the War, provided templates for the development of new housing peripheral to London, and it is against this backdrop that discussion on the New Towns Act took place. Parliamentary discussions \citep{UKParliamentNewTowns} were wide-ranging and touched on several themes that would not be out of place in contemporary discussions. Some participants, such as the Minister of Town and Country Planning, advanced arguments against the expansion of existing towns and in favour of development from scratch:

\begin{quotation}
 Many towns have built new housing estates on the outskirts. These have largely failed in their purpose of providing a better life for their people, and have almost invariably become dormitories consisting of members of one income group, with no community life or civic sense. Today there is a need for additional houses, possibly equal in number again to those built between the wars. Are these to be built on the outskirts of our towns, with the same lack of planning and ill consequences as before? If so, I dread to think what sort of place this still fair land of ours will be in 10 or 15 years time. This is our last chance. Many of the houses now to be built must be carefully located in new self-contained communities, if the existing evils are not to be aggravated.
\end{quotation}

Others, such as Viscount Hinchingbrooke, argued for the redevelopment of existing towns and cities:
\begin{quotation}
 Surely, what we want to do is to remodel our old towns and prevent the gradual invasion of the Countryside by the extension of these towns, leaving behind great masses of derelict property. I am afraid that we may be only too easily exchanging that policy of unplanned ribbon development for a policy of planned cannibalism. What ought to be done is to go to the full extent in remodelling existing towns. Many of my hon. Friends represent London constituencies\ldots and we know that their problem is appalling. They are faced with the most acute housing shortage, and they sense the admirable desire of the people for better conditions. Cannot it be done within the framework of the Greater London Plan by developing blitzed sites, by moving the people from the houses they now inhabit to the blitzed sites while remodelling their own houses, leaving the parkways and open spaces proposed by Professor Abercrombie? My point is that the whole thing can be done by remodelling the Greater London area without any extensions of the number of new towns.
\end{quotation}

Themes range from congestion; to Plato and oligarchies; to the need for a diversity of industry for sufficient employment and resilience; to the need for intermingled societal classes and how the \emph{``friendliness and neighbourliness of the slums is lost when displacing inhabitants to new estates''}; to \emph{``golf courses made available for all''}; and, even, beheadings:

\begin{quotation}
 My researches on new towns go back to the time of Sir Thomas More. He was the first person I have discovered to deplore the ``suburban sprawl'' and in his ``Utopia'' there are 54 new towns, each 23 miles apart. Each town is divided into four neighbourhoods, each neighbourhood being laid out with its local centre and community feed centre. Incidentally, Sir Thomas More was beheaded, but that must not be regarded as a precedent for the treatment of town planners\ldots
\end{quotation}

The New Towns Act of 1946, and interventions since, mandated the creation of officially designated `New Towns' for urban growth and expansion and were to be overseen by development corporations. Though idealised as `greenfield' development, numerous instances entailed the expansion of existing towns, though in either case tended towards lower density single-family development and the separation of land-uses. The majority of New Towns took root from the 1940s through the 1960s, and information about their formation can be found interspersed across numerous issues of official publications such as \emph{The London Gazette} \citep{TheGazette}, though more conveniently collated lists (including references to respective London Gazette issues) can also be found \citep{NewTownsWikipedia}.

Subsequent discussion will here refer to `artificial' urban archetypes and `historical' equivalents. The `artificial' archetype includes `New Towns', though also applies to newer forms of development more broadly, including characteristics in keeping with car-centric, single-family suburban enclave, big-box-store-and-parking-lot morphologies. At risk of confusion, the `historic' archetype does not preclude newer developments if they were to reflect characteristics --- granular, walkable, mixed-use morphologies --- that would not be out of place in historical forms of development.

%% file: content/3_towns_dataset.tex
\section{Overview of the New Towns dataset}

The data used for training the artificial town classifiers is based on a subset of locations drawn from a larger dataset spanning 931 towns and cities in England and Wales. The boundaries are identified using a network percolation method based on road intersections, as developed by \citet{Arcaute2016}, after which the census population statistics are aggregated to each boundary and locations with fewer than 5,000 inhabitants are subsequently discarded. In a later step, all New Town and new-town-like locations are identified, and the subsequent analysis proceeds by further discarding all locations with populations larger or smaller than this band of town sizes. This dataset is here introduced by way of some high-level observations.

Town areas increase sublinearly in relation to population sizes \citep[compare][p.41]{Batty2013}\footnote{
  Ordinary least squares (OLS) regression on log-transformed variables results in a power-law scaling parameter of $\alpha=0.87$. However, non-linear least-squares curve-fitting (Levenberg-Marquardt algorithm) applied to untransformed variables suggests a scaling parameter of $\alpha=0.83$. Whereas these parameters provide some sense of the scaling relation between area and population, it is important to note that these values are inherently sensitive to boundary definitions \citep{Cottineau2017}.
} and, in keeping with Zipf's Law, the larger a population, the less frequent\footnote{
  This is often ascribed to a power-law distribution in the form of $p(x)=Cx^{-\alpha}$ \citep[e.g.][]{Newman2006}. Other candidate distributions should be considered before their exclusion \citep{Stumpf2012, Clauset2009}, which is here performed with the aid of the \code{powerlaw} Python package \citep{Alstott2014}. The package automatically derives an $x_{\min}$ by minimising the Kolmogorov-Smirnov test statistic distance between the data and the fit, here deriving $x_{\min}=10,515$. The package includes methods for comparing candidate distributions using loglikelihood ratios, whereby the lognormal can here be identified as superior to the power-law fit. This marginally remains the case when using a higher $x_{\min}=50,000$ threshold (per, \citet{Arcaute2015}). Visual inspection of the Complementary Cumulative Distribution Function (CCDF) in Supplementary Figure~\ref{fig:global_pop_powerlaw} reveals that the data follows the lognormal distribution closely, but only up to populations $\approx 1,000,000$ after which the populations of Leeds, Birmingham, Manchester, and London show divergence towards the longer tail of the power-law fit-line.
}. Densities tend to be higher for larger cities, but this does not preclude smaller towns from having relatively high densities, and ample variation is evident in Supplementary Figure~\ref{global_area_pop_dens}. The definition of city land-area here excludes large encircled undeveloped parcels (such as undeveloped hills or significant green spaces), which would otherwise unduly inflate the area variable.

\begin{figure}[htbp]
  \centering
  \includegraphics[width=\textwidth, keepaspectratio]{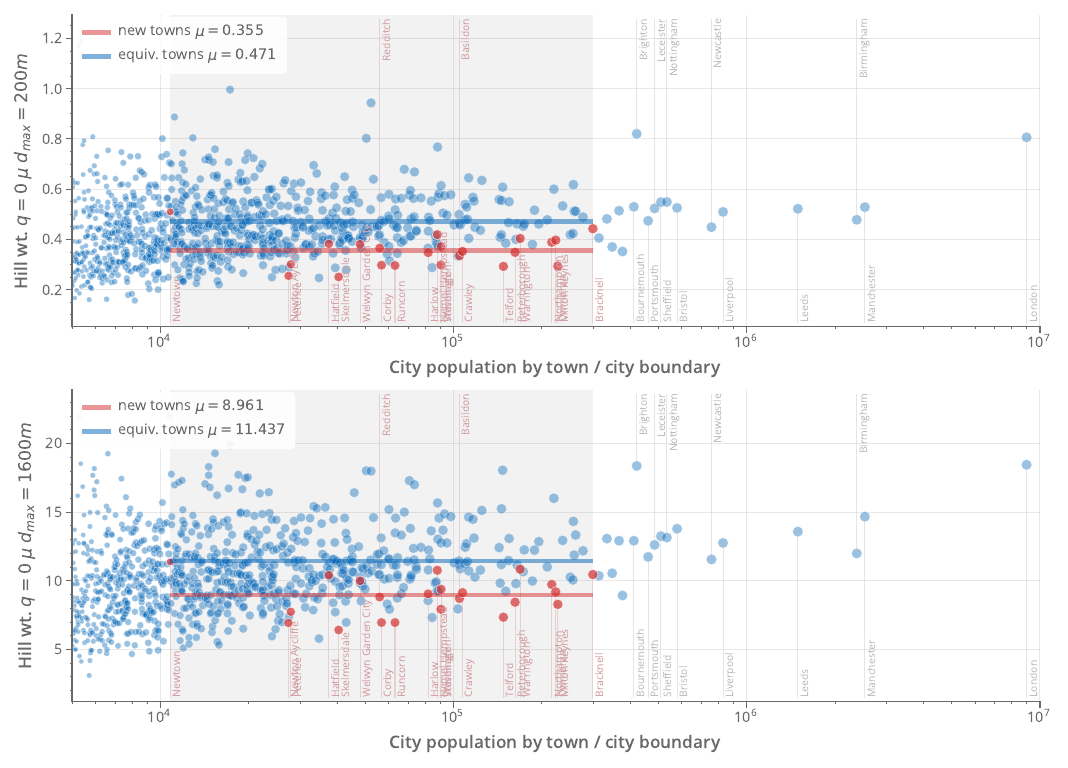}
  \caption[Local mixed-uses by city size]{Local mixed-uses by city size}\label{fig:mus_mu}
 \end{figure}

\begin{figure}[htbp]
  \centering
  \includegraphics[width=\textwidth, keepaspectratio]{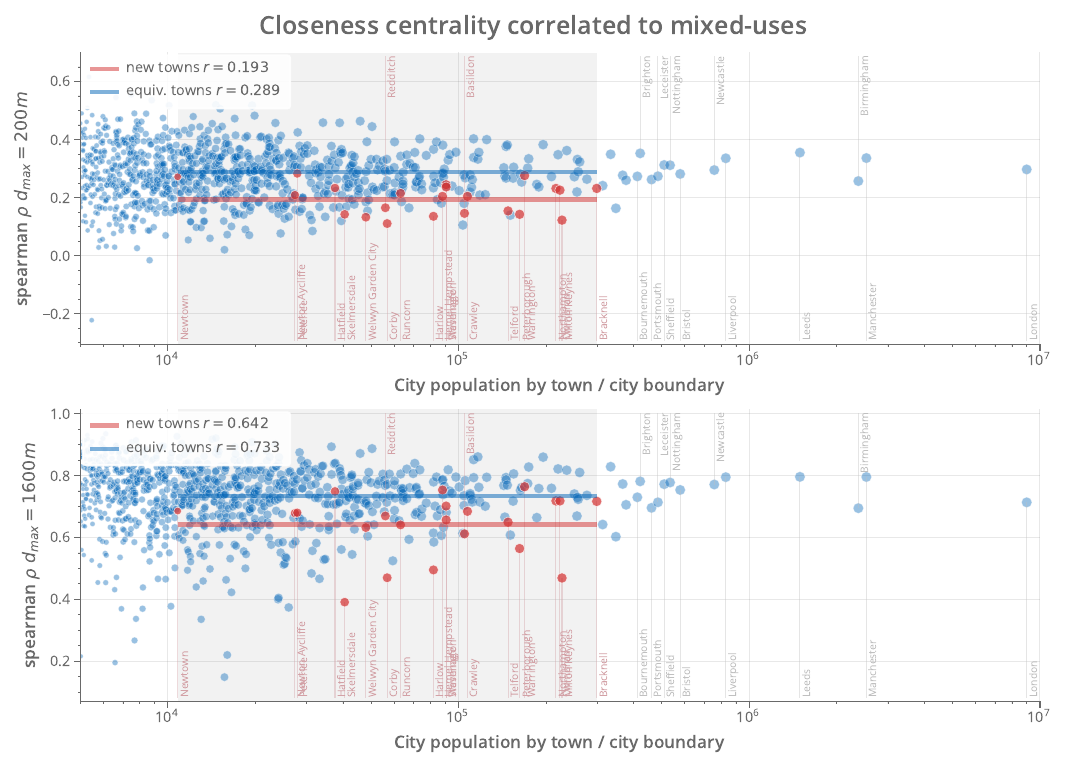}
  \caption[Correlations for local centrality and mixed-uses]{Correlations for local simplest-path harmonic closeness centrality and local mixed-uses by city size}\label{fig:corr_a}
 \end{figure}

Whereas the above provides a brief overview of scaling relationships for town areas and populations, the attention now shifts to an overview of statistics for, specifically, locally computed metrics. Data is sourced from \emph{Ordnance Survey} \emph{Open Roads}, \emph{Ordnance Survey} \emph{Points of Interest}, and \emph{Office for National Statistics} census datasets from which a variety of localised ("radial") centrality, land-use accessibility, mixed-use, and population density measures is derived (see Supplementary Tables~\ref{table:artificial-vars-dnn} and~\ref{table:artificial-vars-m2}). These are computed location-by-location for each street intersection at a selection of pedestrian walking tolerances ranging from $200m$ to $1600m$. This workflow is performed with the open-source \code{cityseer-api} \code{Python} package \citep{Simons2021b}, in which distances are computed directly over the street network from each point of analysis: land-use and statistical aggregations are thus performed dynamically and with respect to the direction of approach towards each land-use or data-point, thereby preserving location-specific relationships between the variables and retaining spatial precision in relation to distance-weighted walking tolerances. Supplementary Figures~\ref{fig:kde_pop} and~\ref{fig:kde_mu} show kernel density estimate plots: larger boundaries tend to have longer tailed distributions and higher median values for local population densities, local closeness centralities, and local mixed-uses. The same variables are averaged and plotted against global city populations in scatterplots Figure~\ref{fig:mus_mu} and Supplementary Figure~\ref{fig:mus_close}, with the mean for bundled values for all New Towns compared to equivalently sized towns. Average local population densities are not substantially different between New Towns and `other' towns, but more notable differences emerge once comparing local closeness centralities and, especially, the degree of local land-use diversity (mixed-uses). This pattern repeats with lower local accessibilities to land-uses in general. The differences likewise persist when comparing the correlations between the variables, as shown in Figure~\ref{fig:corr_a} and Supplementary Figure~\ref{fig:corr_b}, indicating a weakening of the associations between population densities, network centralities, and mixed-uses for New Towns.

%% file: content/4_new_towns_classifiers.tex
\section{Detection of artificial towns with machine learning models}

\begin{figure}[htbp]
 \centering
 \begin{subfigure}{0.45\textwidth}
 \centering\captionsetup{width=0.9\textwidth, justification=centering}
 \includegraphics[width=\textwidth, keepaspectratio]{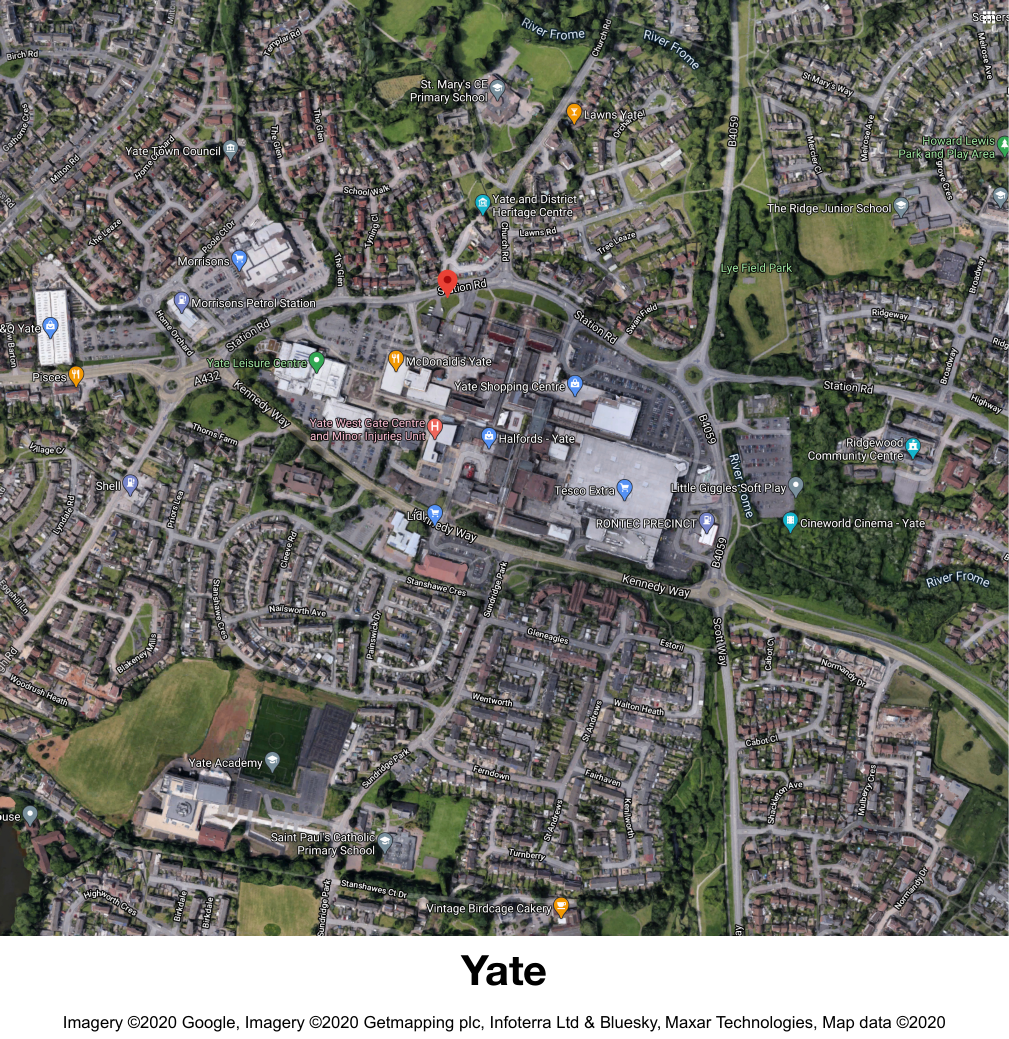}
 \caption[Town of Yate.]{Although not officially designated as a New Town, Yate displays many features consistent with artificial forms of planned development.}\label{fig:yate}
 \end{subfigure}
 \begin{subfigure}{0.45\textwidth}
 \centering\captionsetup{width=0.9\textwidth, justification=centering}
 \includegraphics[width=\textwidth, keepaspectratio]{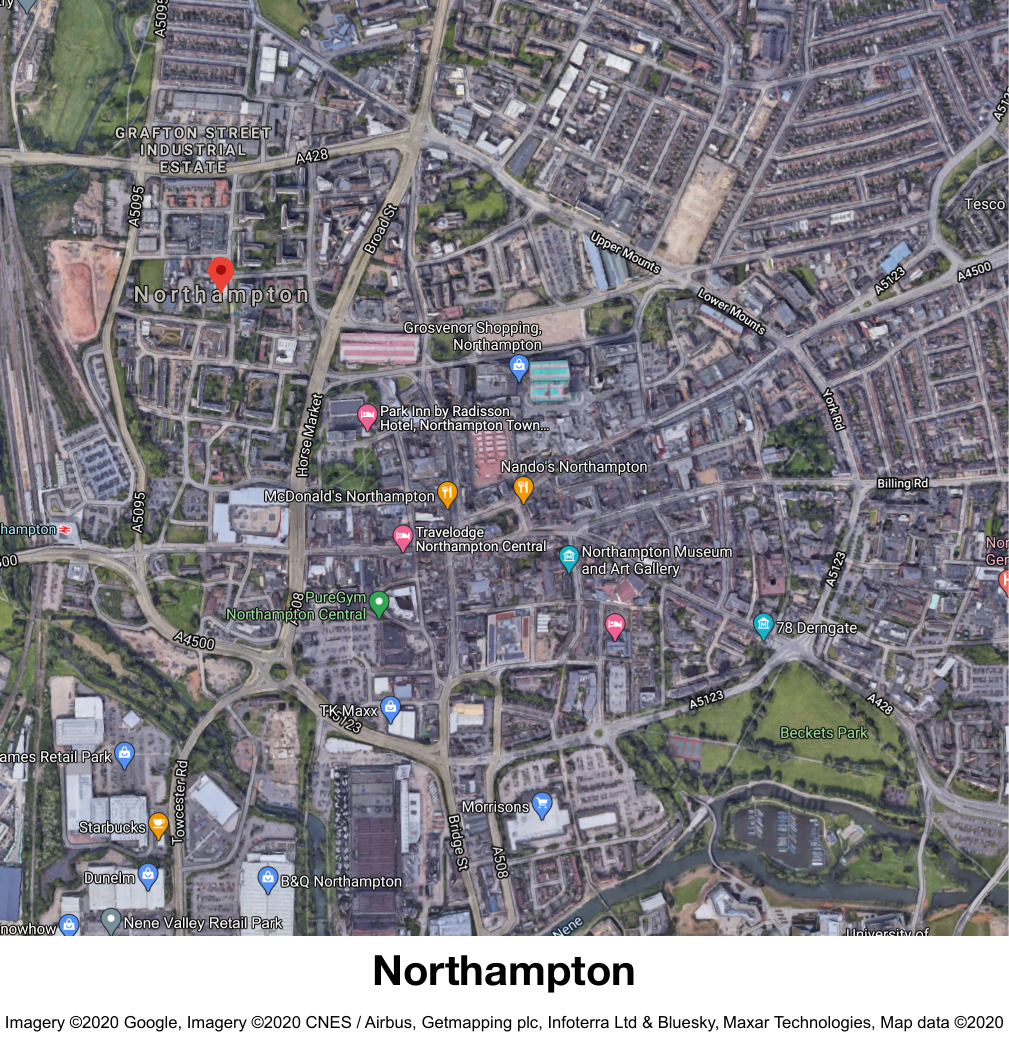}
 \caption[Town of Northampton.]{Designated a New Town in 1968, fragments of the town centre display characteristics consistent with earlier historical development.}\label{fig:northampton}
 \end{subfigure}
 \caption[Discerning historic from artificial forms of development.]{Discerning historic from artificial forms of development.}
\end{figure}

Thus far, the plots shown are reliant on the official designations, in this case, 22 New Town labels. The starting dataset also contains an additional 27 locations that have been identified as `Expanded Towns'. Whereas the respective designations may in some cases be clear-cut, the interpretations thereof are not necessarily so. This conundrum can be illustrated with the use of some examples. The town of Yate (Figure~\ref{fig:yate}), though not designated as a New Town, is characteristic of artificial forms of planned development and features a shopping mall surrounding by parking lots, an automobile-centric roadway system, and clusters of residential enclaves. Conversely, the town of Northampton (Figure~\ref{fig:northampton}), though designated as a New Town in 1968, retains parts of a historic core that had existed well before this time.

\subsection{Fleshing-out town boundary designations}\label{extra-trees-model}

Potential inconsistencies between the character of the respective towns and their official designations and large degrees of variation within respective town boundaries mean that New Town designations offer an incomplete starting point for teaching machine learning classifiers how to distinguish between artificial and historic urban archetypes. Therefore, the first step is to flesh out the dataset by identifying the bulk of other locations that have new-town-like or expanded-town-like characteristics. The visual discrimination of artificial from historical development is a relatively trivial exercise with the benefit of domain expertise, though this faces two problems: it is, firstly, a tedious process in the case of reviewing 931 locations; secondly, it is challenging to comparatively review a large number of locations without a yardstick against which these locations can be consistently compared. For these reasons, a machine learning algorithm is used as a `sidekick' to speed up the identification of candidate locations, which are then inspected and classified accordingly.

As shown in the preceding plots, there is a tendency for New Towns to lag other towns when comparing pedestrian-scale metrics such as closeness centralities, land-use accessibilities, and mixed-use diversity. These indicators are averaged for each town boundary and, complete with existing classifications (i.e. New Town and expanded-town designations), is fed to an Extra-Trees machine-learning classifier\footnote{
  The classification task is performed with a \code{Scikit-learn} Extra-Trees classifier \citep{Pedregosa2011}, selected for its (computational) speed and classification performance for the given dataset and task at hand. Several other classifiers, particularly Random Forests, were reviewed and showed similar (if slower) performance and would likewise be suitable for the current analysis step. Extra Trees classifiers consist of an ensemble of decision trees which filter data through binary decision steps to predict an outcome: starting from a single node, the classifier consequently builds a tree-like structure capable of filtering data into distinct bundles of classifications. Decision trees are grown using algorithms employing one of several techniques, though the intuition is that each node seeks a splitting point for a given feature that would subsequently reduce the `impurity' of the data passing through that step; for example, Gini impurity or cross-entropy for classification tasks, or mean-squared-error for regression tasks. Despite their simplicity, decision trees can fit complex non-linear data but suffer a tendency to overfit and can produce notoriously noisy results. They consequently function better in ensembles --- such as Extra Trees --- which reduce variance by averaging between collections of the inherently noisy, but low bias, decision trees \citep{Hastie2013, Geron2017, Raschka2016}. Hyperparameters were selected using a cross-validated grid-search, giving a final Extra Trees ensemble of 500 estimators using the entropy criterion and a maximum tree depth of 10.
} tasked with learning and subsequently discriminating artificial forms of development from historic. Each iteration proceeds as follows:
\begin{itemize}
  \item The dataset is divided using an 80\%/20\% train/test split, with the same split used for subsequent iterations;
  \item The training set is fed to the model which learns to discern `artificial towns' based on thus-far already labelled examples of New Towns and expanded-towns;
  \item After each training step, the model is applied to the dataset and predicts artificial from historic locations. 
  \item Because the model is being trained on an impartially labelled dataset, it initially predicts a substantial number of `false positive' candidate locations exhibiting characteristics that are similar to officially designated New Towns and expanded-towns;
  \item False positive locations are reviewed with the benefit of domain expertise and are classified accordingly;
  \item The process repeats until the `human-in-the-loop' no longer identifies additional locations for labelling.
\end{itemize}
Earlier iterations of the model identify the more extreme cases of artificial development whereas later iterations increasingly flag locations such as historic cores that were later surrounded to varying degrees by suburban sprawl. In reality, few historic locations have not been affected to some degree, and there is consequently no clear cut-off at which towns are either primarily artificial or historic: the determination of this threshold is left to human judgement and the model aligns accordingly. This process leads to the identification of 185 towns listed in Figure~\ref{text:artificial_towns} consisting of either artificial development or a sufficiently substantial amount of artificial development that has been added to previously `historic' cores. The final iteration of the model attains a 96\%/89\% train/test f1-score (weighted average of precision and recall)\footnote{
  The test-set is not used for model training and in the conventional sense this affords a measure of rigour when evaluating model performance if generalised to unseen data. In this case, the accuracy metric assumes a different nuance because the classifications are iteratively developed using a closed-loop dataset under human supervision, meaning that the model is not being generalised to other datasets and the accuracy metric is, in essence, simply a reflection of the extent to which the model is capable of emulating the human-in-the-loop's discriminatory threshold for discerning artificial from historic forms of development.
}.

\begin{figure}[htbp]
  \centering
  \noindent
  \fbox{
    \begin{minipage}{\textwidth}
    \emph{Abingdon, Addlestone, Amesbury, Andover, Ashford, Aylesbury, Banbury, Bargoed, Barry \& Gibbons Down, Basildon, Basingstoke, Baughurst, Belah, Bexhill, Bicester, Bishops Cleeve, Bodmin, Bognor Regis, Bordon, Brackley, Bracknell, Braintree, Brandon \& Meadowfield, Bridgend, Broadmeadows, Broadway \& Littlemoor, Bromsgrove, Brough, Burgess Hill, Burnley, Burntwood, Bury St Edmunds, Calne, Cambourne, Cambourne \& Redruth, Canvey Island, Carcroft, Carterton, Castleford \& Normanton, Chadwell St Mary, Chapeltown, Chippenham, Cinderford, Clacton On Sea, Clevedon, Coalville, Copperhouse, Corby, Cramlington, Crawley, Cwmbran, Daventry, Deeside, Didcot, Downside, Droitwich, Dunmow, Eastleigh, Eaton Ford, Ellesmere Port, Euxton, Fair Oak, Farnborough, Farncombe, Felixstowe, Gainsborough, Gloucester, Grantham, Grays, Hadfield \& Brookfield, Hailsham, Hammonds Green, Harlow, Harwich, Hastings (\& Bexhill), Hatfield, Haverhill, Haywards Heath, Hemel Hempstead, High Wycombe, Hipswell, Honiton, Horley, Horsham, Hoyland, Huntingdon, Keynsham, Kidderminster, Kidlington, Killamarsh, Kings Lynn, Kingsteignton, Kirkby, Larkfield, Leabrooks, Lee-on-the-Solent, Leighton Buzzard, Letchworth (Garden City), Leyland, Littlehampton, Llandudno Junction, Locks Heath, Lowton, Luton, Maghull, Melksham, Mid Norfolk, Mildenhall, Milton Keynes, Moons Moat South, Morton, Mount Sorrel \& Environs, New Addington, Newbury, Newhaven Town, Newton Aycliffe, Northampton, Northwich, Nuneaton, Peterborough, Peterlee, Plymstock, Port Talbot \& Neath, Primethorpe \& Broughton Astley, Raunds, Redditch, Rochford, Rotherham, Rugeley, Runcorn, Rushden, Ryton \& Crawcrook, Sandy, Seaton Delaval, Shepton Mallet, Shirehampton \& Avonmouth, Sittingbourne, Skelmersdale, Slough, Snodland, South Woodham Ferrers, Southwater, St Neots, Stafford, Stanford-le-Hope, Stevenage, Stocksbridge, Stowmarket, Sudbury, Swadlincote, Swallownest, Swanley, Swindon, Tamworth, Taunton, Taverham, Telford, Thetford, Tidworth, Tiverton, Tonbridge, Trowbridge, Vickerstown, Warrington, Washington, Wath Upon Dearne \& environs, Wellingborough, Welwyn Garden City, West Ravendale, Westwells \& Pickwick, Widnes, Winsford, Winshill, Witham, Wombourne, Wymondham, Yate, Yeovil}
  \end{minipage}
  }
  \caption{List of locations deemed as predominately artificial development}\label{text:artificial_towns}
 \end{figure}

\subsection{Tapping into the full resolution data with a deep neural network}\label{dnn-model}

The first step, which resorted only to pedestrian-scale metrics averaged to town boundaries, is to a degree sufficient to distinguish artificial development from historical in a sense that is arguably consistent with human observations on a town-by-town basis. However, it lacks spatial specificity, and its utility is limited to data averages per town. Further, as already described, the delineation between artificial and historic archetypes is not entirely clear-cut: even the most `historic' of towns are in reality somewhat mixed, with newer residential enclaves, business parks, and shopping malls having been added to varying degrees. It would therefore be preferable to apply the analysis directly at the scale of localised points of analysis so that variations between artificial and historic archetypes can be explored within a given town. It would, furthermore, be advantageous to adopt a more nuanced perspective dropping the binary yes/no designations of `artificialness' in favour of probabilities indicating the degree to which a particular location may reflect different morphologies.

Accordingly, this step resorts to a full resolution dataset where no averaging of values has taken place, i.e. the raw data is fed directly to the ML classifier and is evaluated for each distinct point of analysis. These points are sampled at $20m$ intervals on the road network and the same centrality, land-use, mixed-use, and population metrics (see Supplementary Table~\ref{table:artificial-vars-dnn}) are computed for a range of pedestrian distances relative to each point of analysis. Data is selected from towns with populations falling within the population band of artificial towns identified from the prior step, resulting in a dataset containing approximately 5.2 million points for towns and cities in England and Wales. As a starting point, all points located within artificial town boundaries are labelled accordingly, and the a ML classifier then attempts to learn a mapping from the input variables to the labelled classification.

The challenge at this step of analysis is that the higher resolution model is being trained against classifications derived from the lower-resolution classifications for an enclosing town boundary, meaning that whereas the formal classifications provide a general signpost they do not necessarily faithfully represent the situation for each point of analysis. This strategy relies on the the momentum established by the majority of data points as a whole to steer the model in the correct direction, and is combined with regularisation to ensure that the model does not compensate by overfitting the data. The model consequently associates patterns occurring predominately in either artificial or historic categories with a higher degree of certainty, whereas patterns that feature in both situations trigger a more ambiguous outcome (probability).

The classifier used for this step is based on a deep neural net\footnote{
  Neural networks confer several advantages: they tend to outperform linear methods; scale relatively well to large datasets, and do not suffer the sometimes severe performance penalties associated with other non-linear strategies such as polynomial features or kernel-based methods; their architecture is inherently flexible in that the model is free to transform, isolate, and combine themes within the data; and their implementation can be adapted to specific model architectures and workflows when using a library such as Tensorflow \citep{Abadi2016}. A more comprehensive description of neural networks is provided in \emph{withheld for review}. 
} (DNN) with four hidden layers feeding a sigmoid activation output layer producing the probability associated with the degree of artificiality. This strategy is combined with spatially compartmentalised train/validation sets (75\%/25\%) to prevent spatial leakage of information from the training set into the validation set.

\begin{figure}[htbp]
 \centering
 \includegraphics[width=\textwidth, height=0.95\textheight, keepaspectratio]{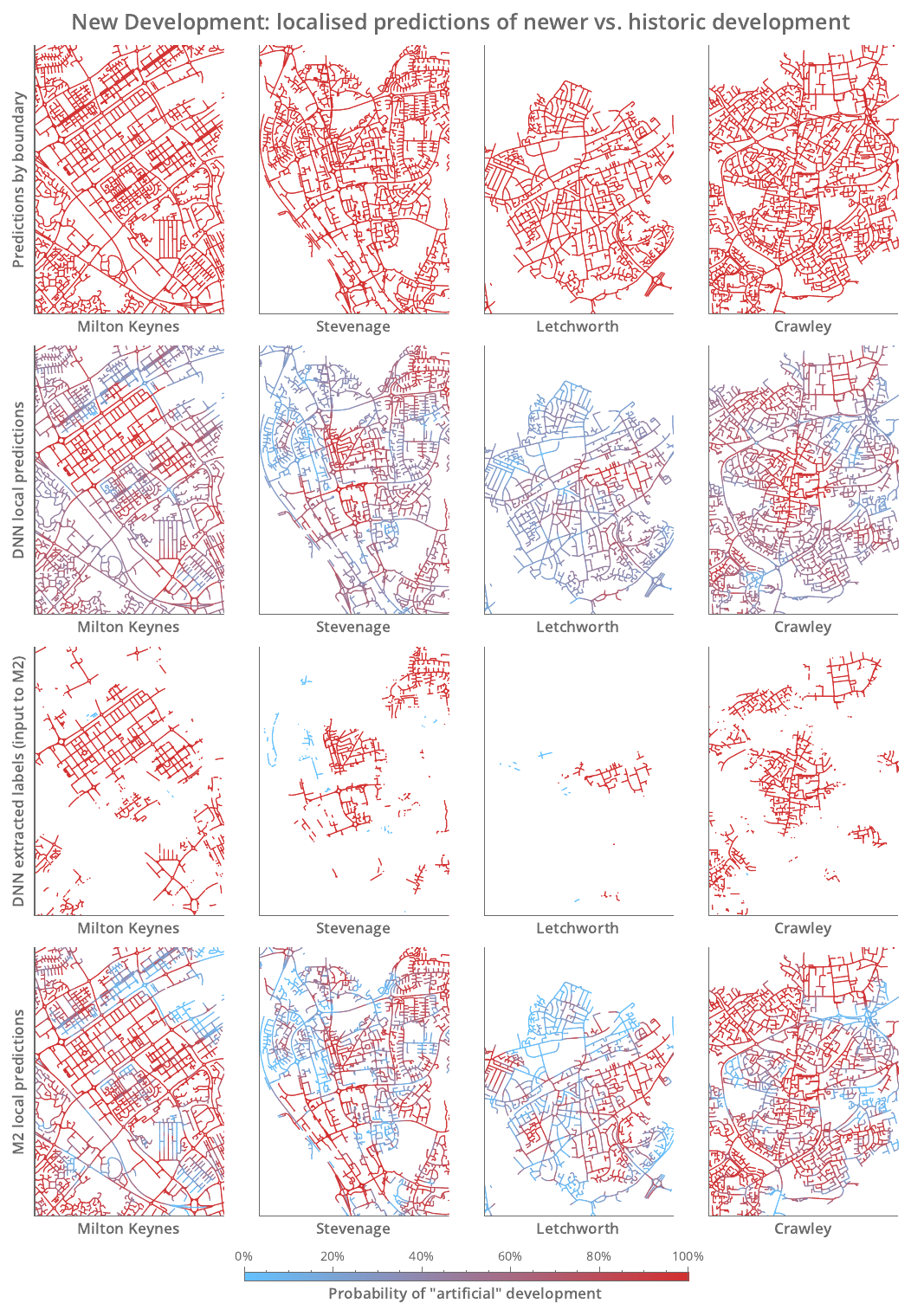}
 \caption[Artificial towns: Comparative predictive methods.]{Comparative predictive methods for selected artificial towns: The top row shows all local points labelled according to the containing town boundary. The second row shows the outcome of the deep neural network, the third-row shows which points are fed to the M2 model in a supervised capacity. The bottom row shows the outputs from the M2 model.}\label{fig:local_pred_artificial}
\end{figure}

\begin{figure}[htbp]
 \centering
 \includegraphics[width=\textwidth, height=0.95\textheight, keepaspectratio]{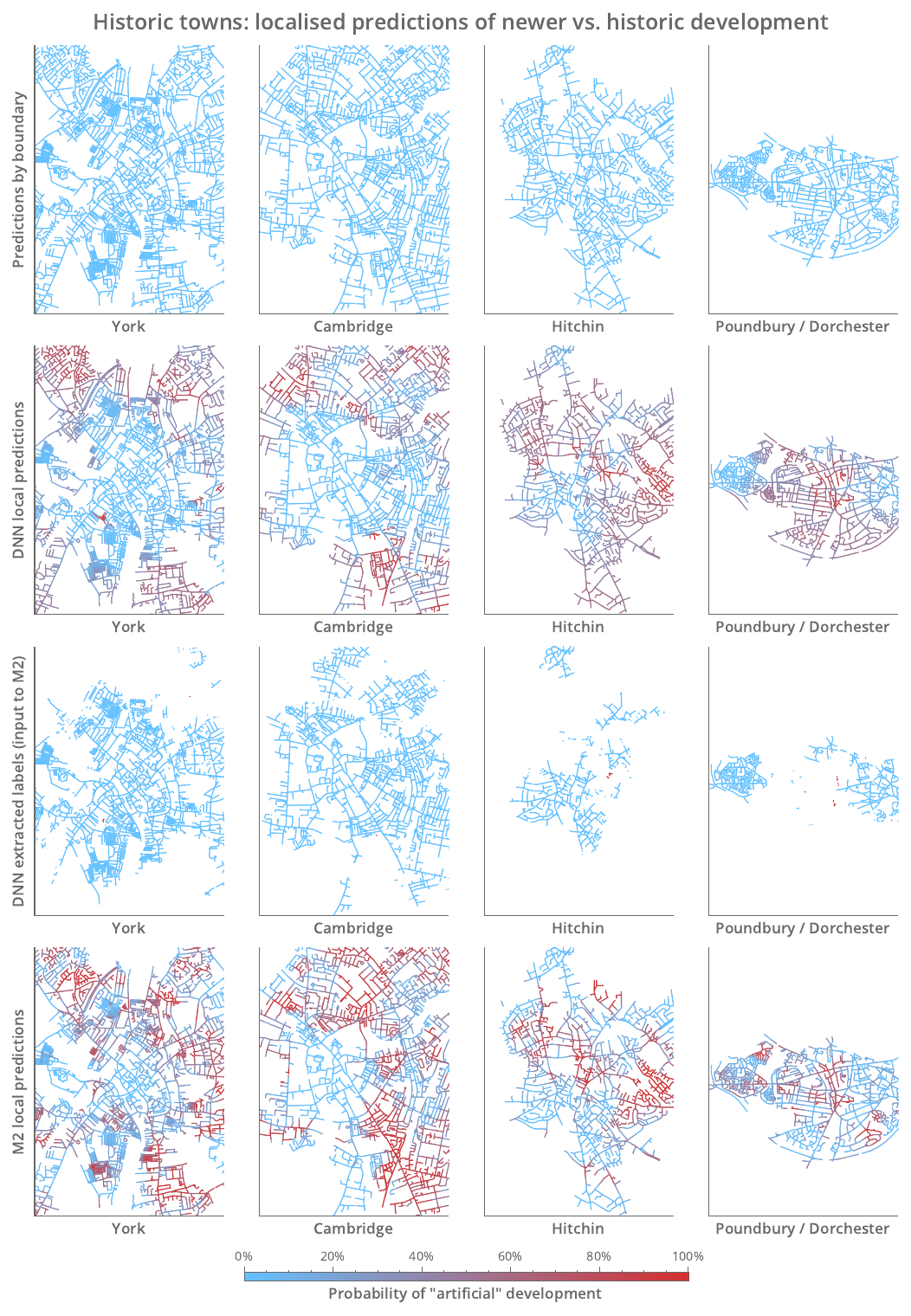}
 \caption[Historic towns: Comparative predictive methods.]{Comparative predictive methods for selected historic towns: The top row shows all local points labelled according to the containing town boundary. The second row shows the outcome of the deep neural network, the third-row shows which points are fed to the M2 model in a supervised capacity. The bottom row shows the outputs from the M2 model.}\label{fig:local_pred_historic}
\end{figure}

Four artificial towns are plotted to Figure~\ref{fig:local_pred_artificial}: Milton Keynes, Stevenage, Letchworth, and Crawley. The first line shows the input classifications derived from the boundaries identified from the human-in-the-loop Extra-Trees classification step; the second line shows the output probabilities from the DNN model. Areas with the most robust probability of being artificial (deep red) predominately include big-box-like districts, car-centric commercial areas, and suburban enclaves with low walkable access to land-uses. Areas with more ambigious probabilities (therefore more likely to overlap with locations found in historic towns) include locations such as pedestrianised high streets or residential locations within walking distances of functional land-uses such as schools or food retail. In this regard, it can be seen that certain planned pedestrian-intensive retail zones such as Letchworth's Leys Ave. are capable of emulating a mix of walkable land-uses in a manner that is not dramatically removed from historic towns.

Four historic towns are likewise plotted to Figure~\ref{fig:local_pred_historic}: York, Cambridge, Hitchin, and Dorchester, the latter of which is included so that the planned development of Poundbury can be compared. In these cases, the more walkable and mixed-use historic cores of York, Cambridge, and Hitchin are associated with higher probabilities of being historic (bright blue). In contrast, newer forms of peripheral development, variously including morphologies such as suburban enclaves, big-box stores, and business parks, veer towards artificial archetypes (red). Poundbury, a planned development that has been designed to emulate specific traditional urban morphological characteristics, shows that newer forms of development can buck the trend by adopting a more granular, mixed-use, and walkable character that resembles historic towns. 

\begin{figure}[htbp]
 \centering
 \includegraphics[width=\textwidth, keepaspectratio]{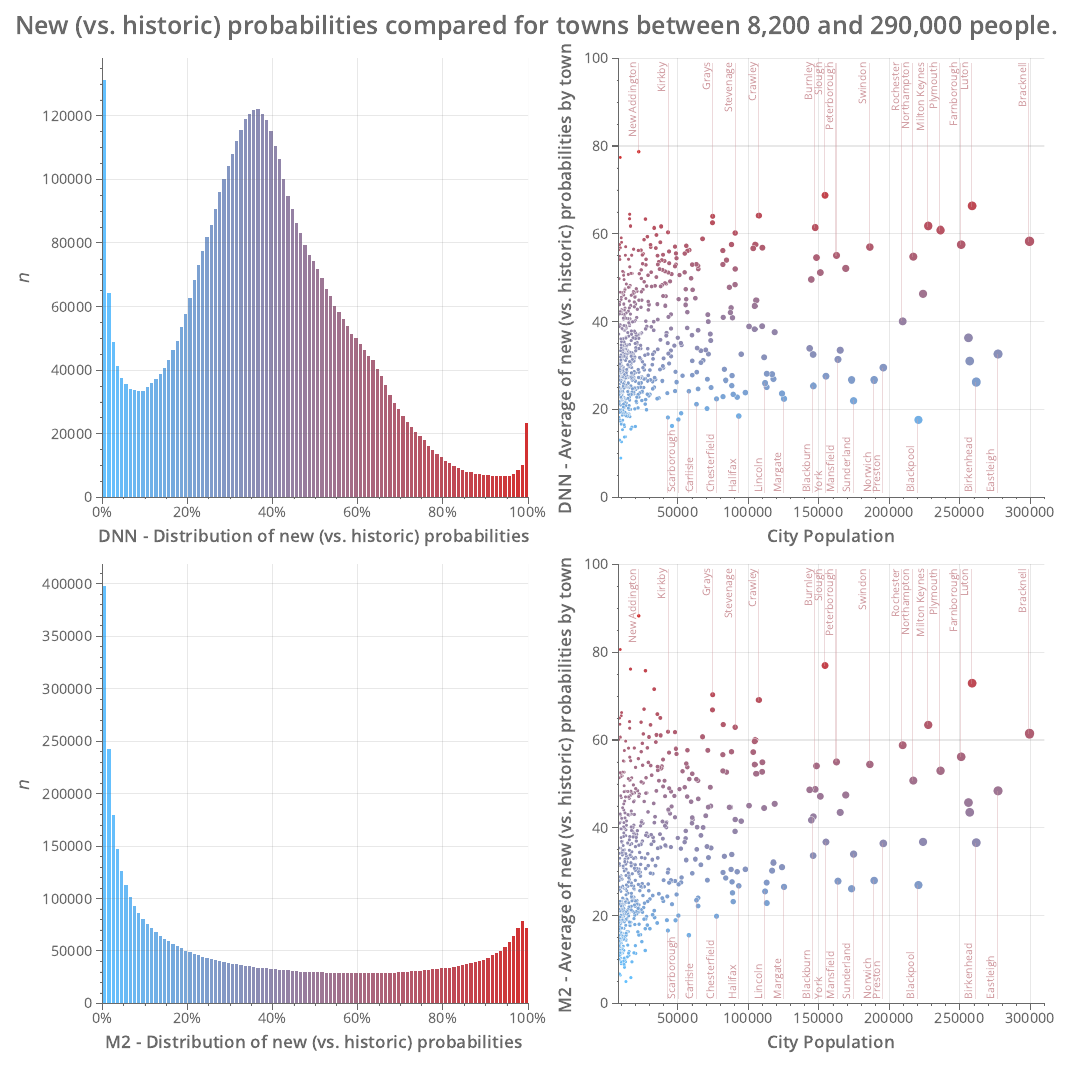}
 \caption[Artificial towns: DNN and M2 distributions and predictions.]{Distributions and predictions for the step 2 Deep Neural Network and the step 3 M2 model.}\label{fig:nt_dists_probs}
\end{figure}

The upper line of Figure~\ref{fig:nt_dists_probs} shows the distribution of probabilities and a scatterplot of probabilities aggregated by town for the DNN model. As may be expected, towns such as Grays, Stevenage, Crawley, Burnley, Slough, Milton Keynes, Luton, and Bracknell take the dubious distinction of being amongst the most artificial. In contrast, older towns such as Scarborough, Chesterfield, Halifax, Margate, York, Sunderland, Norwich, Blackpool, and Birkenhead feature on the opposite end of the spectrum. Interestingly, historical towns with larger populations, such as Blackpool, Birkenhead, and Eastleigh, show clear expanses of planned development; however, unlike the more recent planned developments of the 20th century, these locations demonstrate older planning influences containing enmeshed street networks and assortments of land-uses that have more readily been intermingled with residential. Conversely, as is the case with Cambridge, towns with picturesque historic cores that may have been expected to feature more prominently in the historical category can be held back by newer peripheral development. Examples include business and industrial districts, residential enclaves, and larger-scale forms of institutional development that fragment the urban fabric while leaving little room for granular land-uses to take root.

Whereas the model points in the right direction, there are situations that can be confounding on closer scrutiny. These are likely attributable to three scenarios. The first is the case of murky exemplars, where certain types of newer land-use morphologies such as light industrial or big box stores can feature relatively prominently even in historic towns. The model may therefore deduce that these morphologies are, to a certain degree, not entirely out of place in historical locations. The second situation may be due to the inclusion of larger pedestrian thresholds (up to $1600m$), in which case the model may override impressions given by an immediately adjacent street (e.g. an apparently walkable retail street) due to more distant considerations (such as a lack of nearby population densities). A third situation may be attributable to probability distributions, where potentially ambiguous locations may skew in one direction or the other.

\begin{figure}[htbp]
 \centering
 \includegraphics[width=\textwidth, keepaspectratio]{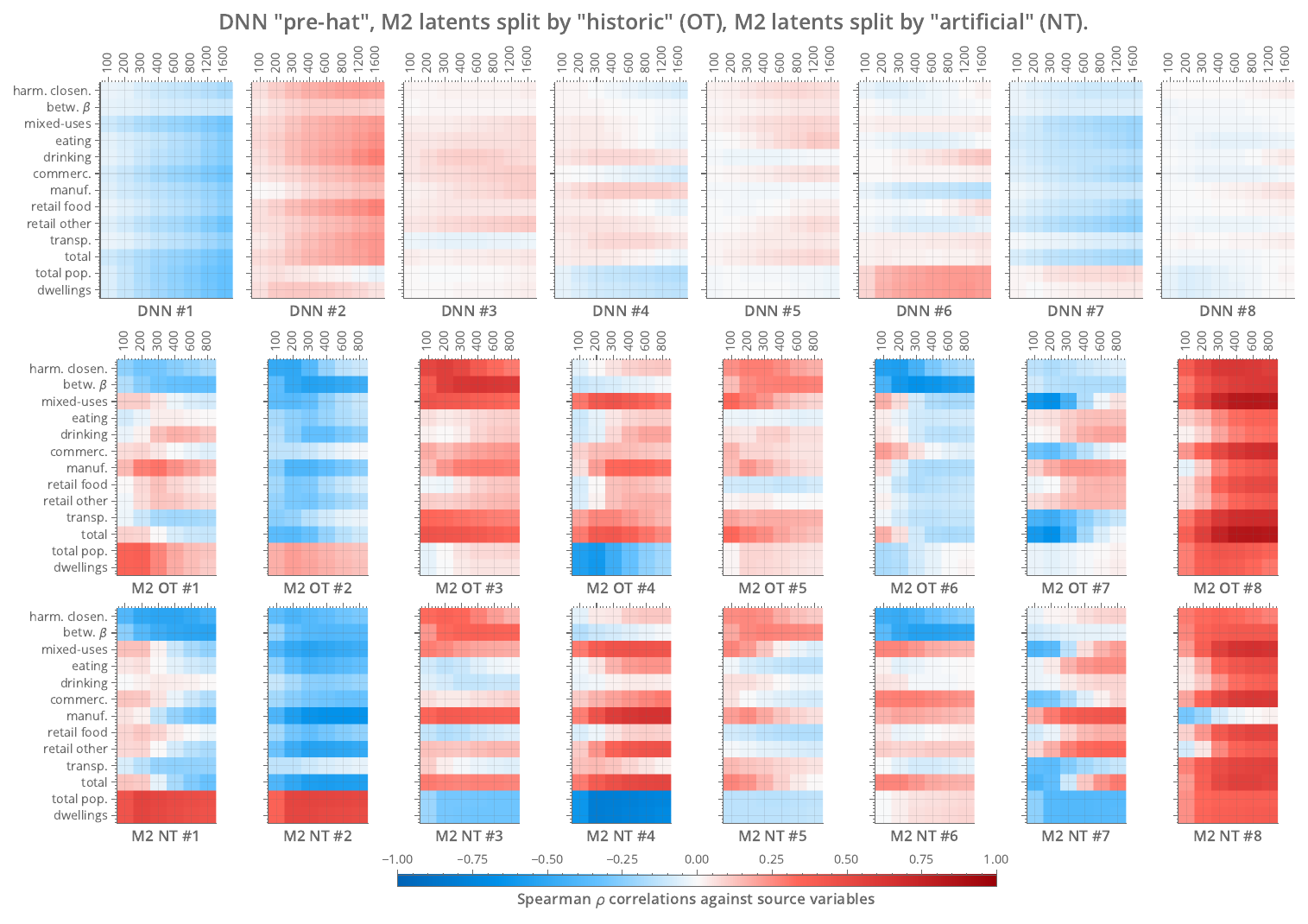}
 \caption[DNN and M2 latents.]{The top row shows `mock latents' of the pre-sigmoid layer for the deep neural network model. The second row shows the M2 latents isolated by artificial locations whereas the third row shows the M2 latents isolated by historic locations.}\label{fig:dnn_m2_latents}
\end{figure}

\subsection{Filling-in the gaps with a semi-supervised model}\label{m2-model}

In the interest of probing the logic within the DNN model, the fourth layer of the neural network (which feeds into the sigmoid output layer) has been reduced to eight neurons, and the (pre-output) state of the model can then be evaluated with respect to the source features. These are shown in the top line of Figure~\ref{fig:dnn_m2_latents} and give an indication of the `themes' and related features corresponding to the outputs of the model. It can be seen that harmonic closeness, mixed-uses, land-use accessibilities, and dwelling densities have a stronger bearing on the model, whereas betweenness centrality, manufacturing land-uses, and transportation have a lesser influence. The model tends to focus on the larger pedestrian thresholds, and it can be theorised that this is a side-effect of the blanket-wise targets imposed on the DNN model (based on town boundaries), thereby encouraging the model to zero-in on wider-area characteristics with lower variances remaining more consistently aligned with a town's overarching classification --- by implication, the model pays less attention to more localised features with higher statistical variances.

In an attempt to release the model from the `tension' induced by the overarching town labels, a semi-supervised \emph{M2} model is here considered as an additional step with the aim of recovering more local specificity. The \emph{M2} model was proposed by \citet{Kingma2014} as a semi-supervised method intended for applications that have access to large quantities of data wherein only a limited number of samples are labelled\footnote{
  The M2 model was introduced alongside discussion of the \emph{M1} model, which uses a variational autoencoder to transform the input data into reduced dimensional latent features. Variational autoencoders are discussed at length in \emph{withheld for review}. 
  The M2 model is a generalisation of variational autoencoders to include a latent class variable $y$ from which unknown classes are inferred: it can be used either in isolation or stacked on top of the outputs of the M1 model, in which case the data is first transformed into a lower-dimensional feature space using M1 and is then fed to the M2 model. Although the authors found that stacked models offered the best performance for their use-case, for the dataset at hand, it was found that unsupervised dimensionality reduction resulted in feature spaces that did not necessarily improve the M2 model's ability to discriminate between the target classes; the M2 model is therefore applied directly to the source variables. Joint optimisation of a reduced dimensionality subspace in concert with class inference was also considered, in the spirit of the VaDE model discussed in \emph{withheld for review}, but did not, in this instance, move beyond tentative development.
}. Samples with the most confident predictions from the DNN model are now used as labelled inputs to the M2 model, meaning that the very highest (artificial) and very lowest (historic) probability-based classifications are retained while the more ambiguous labels are discarded\footnote{
  The upper and lower probability thresholds are set at 25\%, and the larger class (historical locations) is then randomly downsampled so that the classes are balanced (which prevents the model from being pulled excessively in one direction). The result is 5\% of samples in each of the artificial and historical classes for a combined total of 10\% labelled samples.
}. In this sense, the model is `supervised': it is anchored at either extreme by locations that are more confidently historic or `artificial', as indicated by the third rows of Figures~\ref{fig:local_pred_artificial} and~\ref{fig:local_pred_historic}. The remaining 90\% of labels is discarded, and the M2 model is then free to extrapolate the natural patterns and distributions within the data (see Supplementary Table~\ref{table:artificial-vars-m2}), and it is in this sense that the model is `unsupervised'.

The scatterplot (Figure~\ref{fig:nt_dists_probs}) of M2 probabilities by town shows a relatively similar combined outcome to the DNN model, though with subtle changes to the ordering of locations. The lower line of figures within, respectively, Figures~\ref{fig:local_pred_artificial} and~\ref{fig:local_pred_historic}, show that the M2 model adopts a more opinionated disposition and has a tendency to pull the probabilities towards either extreme. Part of the reason for considering the M2 model is that the model's latent features can be explored per Figure~\ref{fig:dnn_m2_latents}. Whereas the interpretation of latents is complex (because they can vary and interact across multiple dimensions), it remains possible to tease out subtle distinctions by comparing correlations for latents split by artificial instead of historic locations. Historic locations show more substantial and consistent collinearity between closeness centralities, land-uses, and densities (latent \#8); artificial locations, in contrast, show more pronounced shear between all of the major classes of variables: between network centralities and land-uses (latents \#1, \#6); network centralities and population (latent \#1, \#3); and between densities and land-uses (latents \#1, \#2, \#3, \#4, \#5, \#7). Historic locations place greater emphasis on local land-uses (inverses of latents \#2 and \#7) in contrast to artificial locations, which place greater emphasis on manufacturing land-uses through all latents, with manufacturing land-uses tending to be more clearly separated from other land-uses when compared to historical locations (latent \#8).

%% file: content/5_summary.tex
\section{Summary}

Machine learning methods are tremendously powerful, but their capabilities are often oversold: they can be dangerous if used without sufficient oversight and, as with statistics and mathematical modelling more generally, can be misused. On the other hand, planners, policymakers, and politicians have struggled to translate the need for walkable and diverse urbanism into a tangible form, and there tends to be a disconnect between good intentions and oft defunct artificially planned communities. It is here argued that supervised and semi-supervised machine learning models --- if developed with rigorous input and oversight from domain experts --- offers the opportunity to explore scalable workflows that, on the one hand, align with the instincts of urbanists and, on the other, help to better convey the ramifications of planning decisions.

A worked example shows how that strictly pedestrian-scale information derived from the road network, the mix and accessibility of land-uses, and the density of dwellings and populations can be linked to different types of machine learning models in support of exploring spatial distinctions between artificial and historical urban archetypes. These examples are not to be construed as optimal but are here provided as `proof-of-concepts' that can be further improved through more extensive exploration and testing tailored to available sources of data and the envisioned end-goals. For emphasis, such tools are not intended to replace but rather reinforce existing design specialities by making it possible to better gauge the potential outcomes of decisions at scale. Further, these are not envisioned to replace existing review and approvals processes but rather to bolster these by providing robust and repeatable methods that can help to improve the transparency and accountability of decision-making processes.

It is encouraging that the models, which have been trained using only the supervised targets derived from artificial or historic towns, have been capable of exploring themes relating of closeness centralities, the mix and accessibility of land-uses, the density of dwellings and populations, and the relationships between these. This behaviour is broadly consistent with intuitions expressed by urbanists and shows that such methods hold potential for further development. This research has elucidated three issues that warrant further investigation in subsequent research: firstly, the difficulty in cultivating clearly delineated exemplars for training the models, and which can likely be resolved with smaller sets of curated exemplars hand-picked by domain experts; secondly, the tendency for models to rely on the lower variances of variables calculated for measures at larger distance thresholds, thus warranting consideration for strategies better emphasising more localised thresholds but not to the exclusion of wider-area information; and thirdly, further development of methods to calibrate the probability distributions generated by the classifiers to concur with the intuition of domain experts for a range of scenarios.

%% file: content/S1.tex
\subsection{Dataset exploratory plots}

\begin{figure}[h]
  \centering
  \includegraphics[width=\textwidth, keepaspectratio]{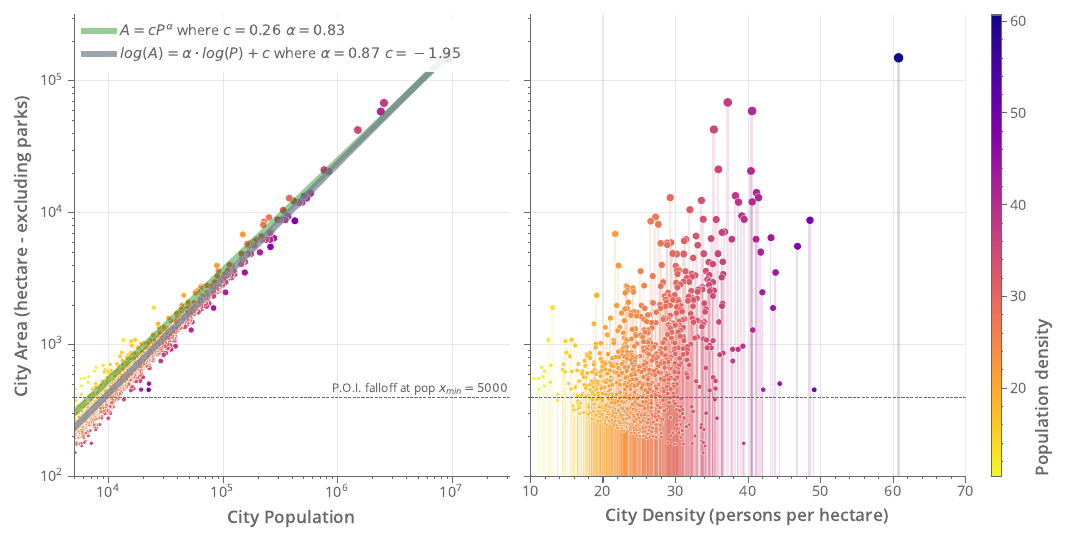}
  \caption[Scatter plots: population, area, and density]{Left: Scatter plot and regression for population and area. Right: Population density}\label{global_area_pop_dens}
 \end{figure}
 
 \begin{figure}[h]
  \centering
  \includegraphics[width=\textwidth, keepaspectratio]{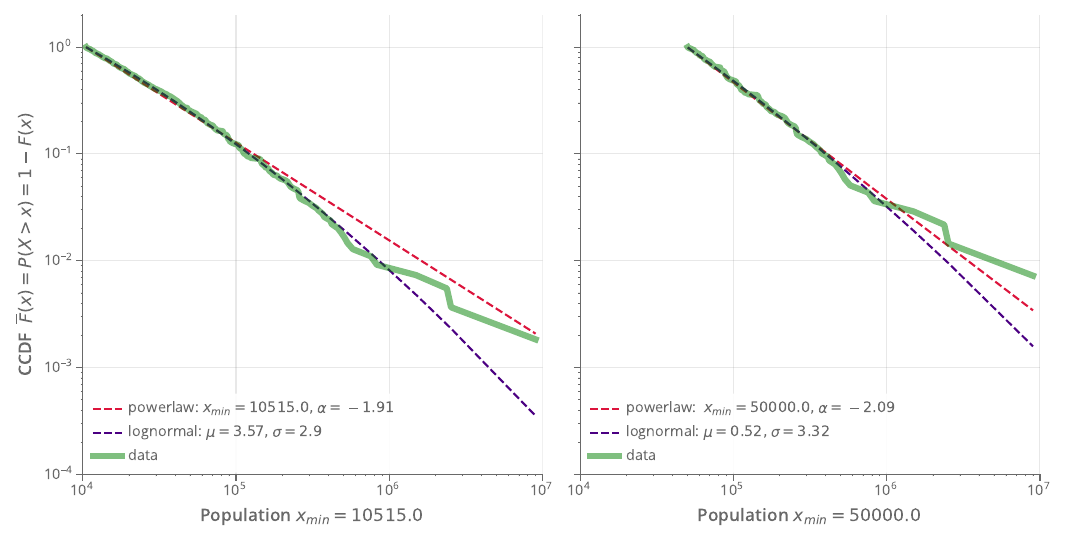}
  \caption[CCDF for city population vs. Lognormal and Powerlaw fit]{Left: CCDF for city population vs. Lognormal and Powerlaw fit with $X_{\min}=10,515$. Right: CCDF for city population vs. Lognormal and Powerlaw fit with $X_{\min}=50,000$}\label{fig:global_pop_powerlaw}
 \end{figure}

\begin{figure}[h]
  \centering
  \includegraphics[width=\textwidth, keepaspectratio]{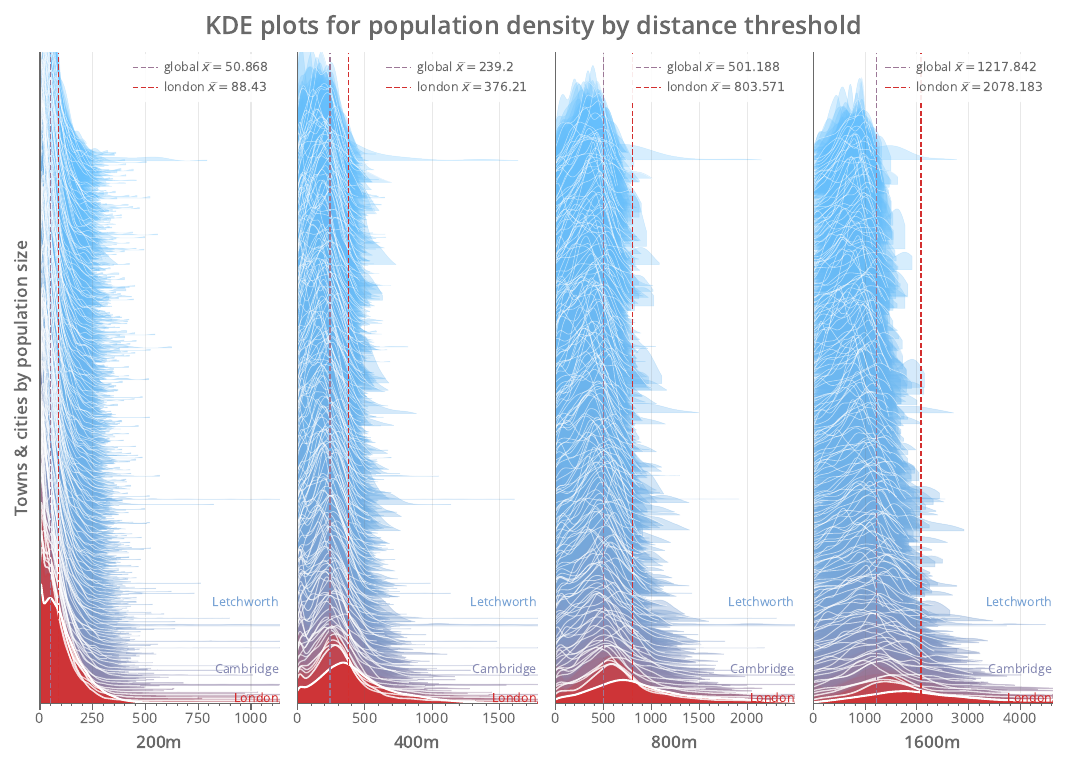}
  \caption[KDE for local population by city size]{KDE for local population densities to all city boundaries from smallest (blue) to largest (red)}\label{fig:kde_pop}
 \end{figure}
 
 \begin{figure}[h]
  \centering
  \includegraphics[width=\textwidth, keepaspectratio]{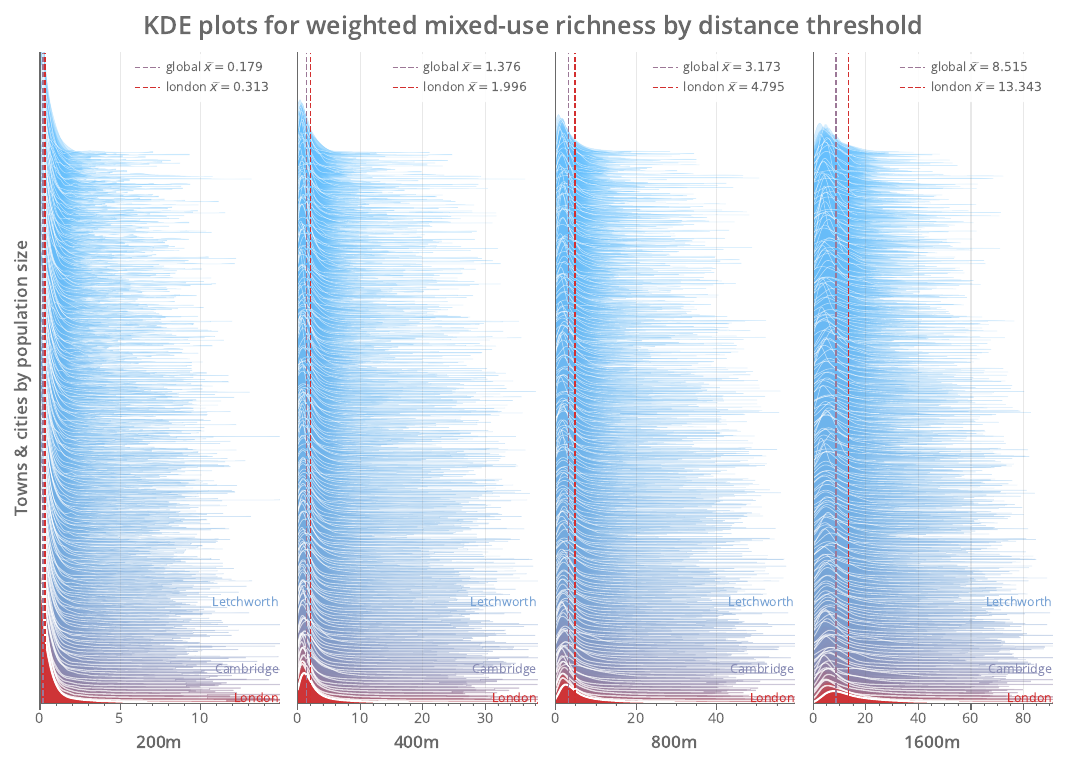}
  \caption[KDE for local mixed-uses by city size]{KDE for local mixed-uses to all city boundaries from smallest (blue) to largest (red)}\label{fig:kde_mu}
 \end{figure}
 
 \begin{figure}[h]
  \centering
  \includegraphics[width=\textwidth, keepaspectratio]{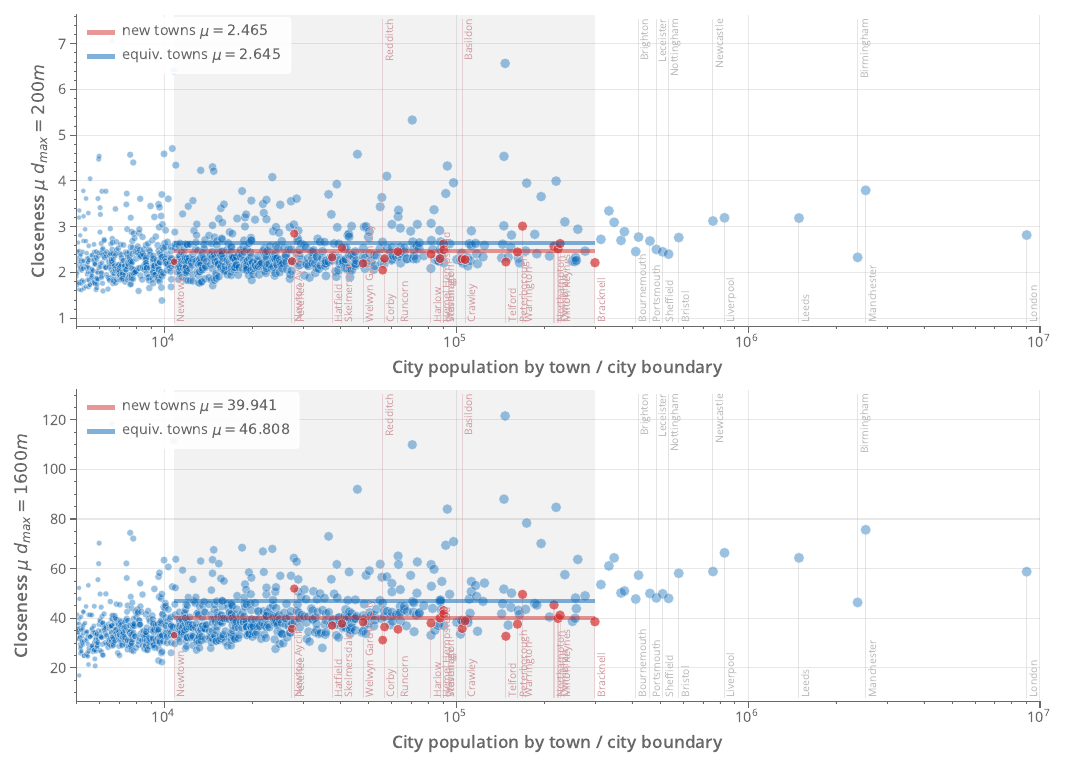}
  \caption[Local closeness by city size]{Local simplest-path harmonic closeness centrality by city size}\label{fig:mus_close}
 \end{figure}
 
 \begin{figure}[h]
  \centering
  \includegraphics[width=\textwidth, keepaspectratio]{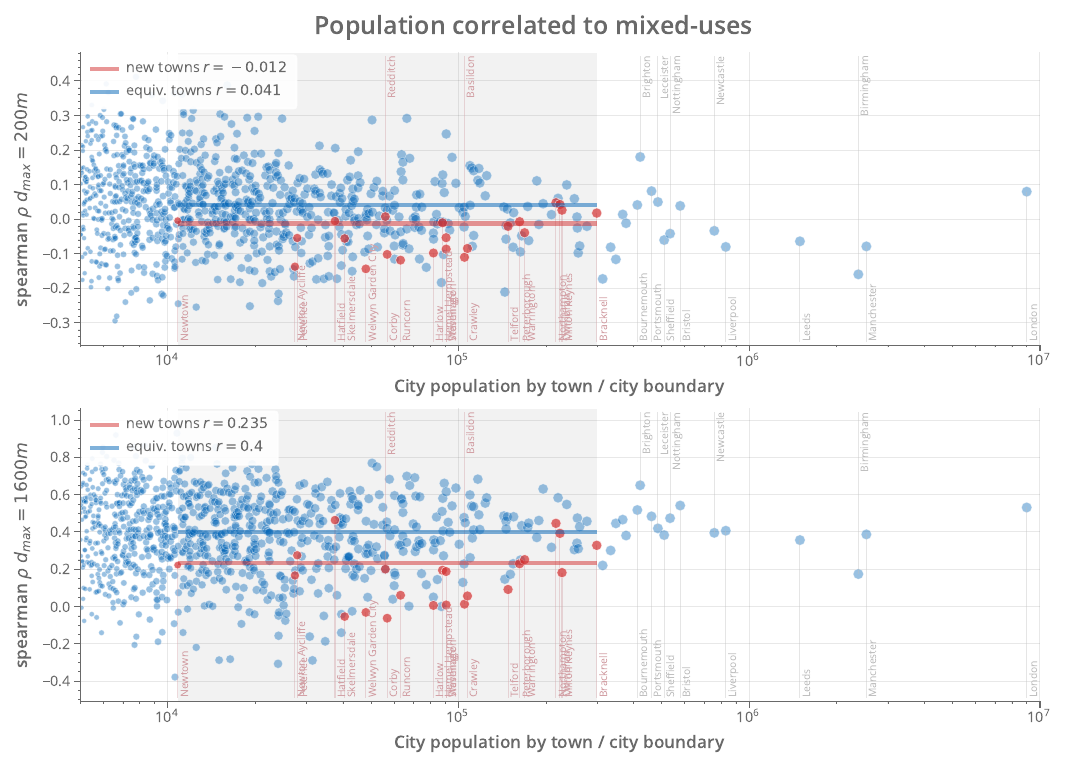}
  \caption[Correlations for local population and mixed-uses]{Correlations for local population density and local mixed-uses by city size}\label{fig:corr_b}
 \end{figure}

\clearpage
\subsection{Variables used for model training.}

\begin{table}[ht]
  \centering\footnotesize
  \begin{tabular}{ p{1.5cm} | p{5cm} p{2cm} p{2cm} }
    &
    Variables
    &
    Distances
    &
    Data Source
    \\
    \midrule
    Centrality data
    &
    Node harmonic closeness centrality\newline
    Node betweenness beta centrality
    &
    100m to 1600m
    &
    OS Open Roads
    \\
    \midrule
    Land-use data
    &
    Mixed-use richness\newline
    Eating establishment accessibility\newline
    Drinking establishment accessibility\newline
    Commercial venue accessibility\newline
    Manufacturing location accessibility\newline
    Food retail shop accessibility\newline
    Other retail shop accessibility\newline
    Transportation or station accessibility\newline
    Total landuse accessibility
    &
    100m to 1600m
    &
    OS POI
    \\
    \midrule
    Statistical Data
    &
    Population density\newline
    Dwelling density
    &
    100m to 1600m
    &
    ONS Census
    \\
  \end{tabular}
  \caption[Summary of source variables used for Sections~\ref{extra-trees-model} \&~\ref{dnn-model}.]{Summary of source variables used for the Extra-Trees model boundary classification (aggregated by town boundary) and Deep Neural Network classifier in Sections~\ref{extra-trees-model} \&~\ref{dnn-model}.}\label{table:artificial-vars-dnn}
 \end{table}
 
 \begin{table}[ht]
  \centering\footnotesize
  \begin{tabular}{ p{1.5cm} | p{5cm} p{2cm} p{2cm} }
    &
    Variables
    &
    Distances
    &
    Data Source
    \\
    \midrule
    Centrality data
    &
    Node harmonic closeness centrality\newline
    Node betweenness beta centrality
    &
    100m to 800m
    &
    OS Open Roads
    \\
    \midrule
    Land-use data
    &
    Mixed-use richness\newline
    Eating establishment accessibility\newline
    Drinking establishment accessibility\newline
    Commercial venue accessibility\newline
    Manufacturing location accessibility\newline
    Food retail shop accessibility\newline
    Other retail shop accessibility\newline
    Transportation or station accessibility\newline
    Total landuse accessibility
    &
    100m to 800m
    &
    OS POI
    \\
    \midrule
    Statistical Data
    &
    Population density\newline
    Dwelling density
    &
    100m to 800m
    &
    ONS Census
    \\
  \end{tabular}
  \caption{Summary of source variables used for the M2 semi-supervised classifier in Section~\ref{m2-model}.}\label{table:artificial-vars-m2}
 \end{table}

%% file: shared/acknowledge_phd.tex
\subsection{PhD}

This paper derives from the author's PhD research at the \emph{Centre for Advanced Spatial Analysis}, \emph{University College London}. The author wishes to acknowledge their PhD supervisors, Prof.~Elsa Arcaute and Prof.~Michael Batty, for their gracious support and feedback throughout the development of this work. The author takes sole responsibility for any oversights or shortcomings contained within this paper.

%% file: shared/acknowledge_data.tex
\subsection{Data}

\begin{flushleft}
The geographical plots and statistical figures in this document have been prepared with use of the following sources of data:\linebreak
\linebreak
\textbf{\emph{Ordnance Survey} \emph{Open Roads}}\linebreak
\emph{Contains OS data © Crown copyright and database right 2021.}\linebreak
\linebreak
\textbf{\emph{Ordnance Survey} \emph{Points of Interest} data}\linebreak
\emph{This material includes data licensed from PointX© Database Right/Copyright 2021.}\linebreak
\emph{Ordnance Survey © Crown Copyright 2021. All rights reserved. Licence number 100034829.}\linebreak
\linebreak
\textbf{\emph{UK Data Service} / \emph{Office for National Statistics} census data}\linebreak
\emph{Contains National Statistics data © Crown copyright and database right 2021.} \linebreak
\emph{Contains OS data © Crown copyright and database right (2021).}\linebreak
\end{flushleft}